\documentclass[11pt]{article}
\usepackage{euler}
\usepackage{ams}
\usepackage{xypic}
\newtheorem{la}{Lemma}

\newtheorem{tm}{Theorem}

\newtheorem{ce}{Conjecture}
\newtheorem{st}{Statement}

\input xy
\xyoption{all}

\begin{document}
\begin{center}{\Large{\bf On the Existence of certain Quantum Algorithms}}\\
\vspace{.5cm}
{Bj\"orn Grohmann}\\\vspace{.2cm}
{\small Universit\"at Karlsruhe, Fakult\"at f\"ur Informatik\\76128 Karlsruhe, Germany}\\
{\small {\tt nn@mhorg.de}}
\end{center}
\vspace{.5cm}
\begin{abstract}
We investigate the question if quantum algorithms exist that compute the maximum
of a set of conjugated elements of a given number field in quantum polynomial time. We will relate the existence of these algorithms for a certain family of number fields to an open conjecture from 
elementary number theory.
\end{abstract}
\section{Introduction}
Let $\Q(\Gamma)/\Q$ be a Galois extension with Galois group G and define
$\Gamma_{\rm max}:=\max_\sigma\{\,|\Gamma^\sigma|\,\}$,
with $\sigma\in G$. We ask if there exists an algorithm that, given some description of $\Gamma$, efficiently computes a $\varphi\in G$, such
that $|\Gamma^\varphi|=\Gamma_{\rm max}$.\\\\
The first problem we encounter in this general setting is that two conjugated elements might not be efficiently distinguishable, i.~e.,
for $\sigma,\varrho\in G$, the difference $||\Gamma^\sigma|-|\Gamma^\varrho||$ may become very small. We will avoid this problem by defining,
for a positive integer $t\in\N$, the set:
\begin{equation}
{\rm MAX\Gamma}_{|G|,t}:=\{\,\sigma\in G\,|\,|\Gamma^\sigma/\Gamma_{\rm max}|>1-1/(\log|G|)^t\,\},
\end{equation}
and ask for an efficient computation of an element $\varphi\in {\rm MAX\Gamma}_{|G|,t}$, say, the number of steps being a polynomial
in $\log|G|$.\\\\
In this article, we will relate the existence of these algorithms for a certain family of number fields to an open conjecture from 
elementary number theory in a sense that either algorithms of this kind exist or the conjecture is true or both.\\\\
To state the conjecture, we introduce the {\em Fermat quotient} $q_p(k)$, where $p$ is an odd prime and $k\in \Z$ with $(k,p)=1$, to be
the smallest integer greater or equal to $0$ that satisfies the equation
\begin{equation}
k^{p-1}\equiv 1+q_p(k)p \,\,{\rm mod}\, p^2.
\end{equation}
Here, we are interested in the number of the first consecutive zeros of this 
quotient, that is
\begin{equation}
\kappa_p:=\min\{\,q\in \N\,|\,q_p(q)\not=0\,\}.
\end{equation}
For a long time this integer was closely related to first case of Fermats Last Theorem, but we will not go into this here (see \cite{grohmannzerosfq} and the references given there).\\\\
It has been shown in \cite{grohmannzerosfq} that
\begin{equation}
\kappa_p\in O(\sqrt p),
\end{equation}
and it is still an open question, whether this bound is tight. The following Conjecture lowers this bound for infinitely many primes:
\vspace{0.5cm}
\begin{ce}\label{cekappap}
For all $\epsilon>0$ there exist infinitely many primes $p$ with $\kappa_p<\epsilon\sqrt p$.
\end{ce}
\vspace{0.5cm}
Now, let $p$ be again an odd prime and $\zeta_{p^2}:=e^{2\pi i/p^2}$ a primitve $p^2$-th root of unity. Further, denote
by $\Q(\Gamma_p)/\Q$ the real subfield of $\Q(\zeta_{p^2})$ of degree $p$, where $\Gamma_p$ is given by
\begin{equation}
\Gamma_p:=\sum_{\sigma^{p-1}=1}\zeta_{p^2}^\sigma,
\end{equation} 
with $\sigma\in G(\Q(\zeta_{p^2})/\Q)\simeq (\Z/p^2\Z)^\times$.\\\\
The aim of this article is to prove the following theorem:
\vspace{0.5cm}
\begin{tm}\label{tmmaxgammabqp}
At least one of the following is true:
\begin{enumerate}
\item For all positive integers $t\in \N$, there exist a constant $c_t$ and a quantum algorithm that, given an odd prime $p$, computes in
$(\log p)^{c_t}$ steps and with a probability close to $1$ an element of the set ${\rm MAX\Gamma}_{p,t}$.
\item For all $\epsilon>0$ there exist infinitely many primes $p$ with $\kappa_p<\epsilon\sqrt p$.
\end{enumerate}
\end{tm}
\vspace{0.5cm}
The paper is organized as follows. In the next section, a quantum algorithm is presented that attempts to compute 
an element of the set ${\rm MAX\Gamma}_{p,t}$ in quantum polynomial time, at least if Conjecture \ref{cekappap} is false. Then, after recalling 
some basic facts from
number theory, we will state the proof of the Theorem.\\\\
\section{The Algorithm}
In the following let $p$ be an odd prime.
To present the algorithm, we define a polynomial time computable function $f:\Z\longrightarrow \Z$ by
\begin{equation}
f(x):=\left\{\begin{array}{ll}
p, & \mbox{if }\,x\equiv 0\,\,{\rm mod}\,p\\ 
q_p(x), & \mbox{else}\\
\end{array}\right.,
\end{equation}
where $q_p(x)$ denotes the Fermat quotient of the integer $x$, defined in the last section.\\\\
(i) For the quantum part of the algorithm, we start with the state
\begin{equation}
\frac{1}{p}\sum_{x=0}^{p^2-1}|x\rangle |f(x)\rangle
\end{equation}
and (ii) apply the Quantum Fourier Transform (\rm QFT) to the first register, which leads to
\begin{equation}
\frac{1}{p^2}\sum_{a,x=0}^{p^2-1}\zeta_{p^2}^{ax}\,|a\rangle |f(x)\rangle.
\end{equation}
(iii) We now measure the system and obtain the state $|a\rangle |s\rangle$ with probability
\begin{equation}\label{probas}
\frac{1}{p^4}\,\bigg|\sum_{f(x)=s}\zeta_{p^2}^{ax}\,\bigg|^2.
\end{equation}
(iv) If $a\not\equiv 0\,\,{\rm mod}\,p$ and $s\not=p$, we note down the smallest nonnegative integer $\sigma^\prime$ that satisfies the equation
\begin{equation}
\sigma^{\prime}\equiv q_p(a)+s\,\,{\rm mod}\,p.
\end{equation}
For a constant $c$, to be specified later, we repeat the whole process $(\log p)^c$ times and output the integer $\sigma^\prime$ that has
occured most frequently (in case of a tie we choose one of the ``leaders" by random).\\\\
In order to obtain the desired element $\sigma\in G_p:=G(\Q(\Gamma_p)/\Q)$, we define the group homomorphism 
$j_p:\Z\longrightarrow G(\Q(\zeta_{p^2})/\Q)$ such that $j_p(a)(\zeta_{p^2}):=\zeta_{p^2}^{1-ap}$ and set 
$\sigma:=j_p(\sigma^\prime)|_{\Q(\Gamma_p)}$.
\section{Analysis}
For the analysis of the algorithm, we will begin with two Lemmata stating the probabilities of the possible outcomes of step (iv) of the
quantum subroutine from the last section:\\\\
\vspace{0.5cm}
\begin{la}
The probability that step {\rm (iv)} produces some integer equals $1-\frac{1}{p}\left(2-\frac{1}{p}\right)$.
\end{la}
\vspace{0.5cm}
{\bf Proof.} Let $|a\rangle |s\rangle$ be the state given after the measurement in step (iii) of the procedure. We will collect the probabilities
for the event $s=p$ or $a=kp$, with $k\in\{0, 1, \dots, p-1\}$:\\\\
$s=p$ and $a\not= kp$: The probability for this event equals $0$, since by definition of the function $f$ the inner sum of
equation (\ref{probas}) equals $\sum_{r=0}^{p-1}\zeta_p^{ar}=0$.\\\\ 
$s=p$ and $a=kp$: Here, the inner sum of equation (\ref{probas}) is equal to $p$ and since there are also $p$ possibilities for the
integer $k$, the probability for this event equals $\frac{1}{p}$.\\\\
$s\not=p$ and $a=0$: In this case, the value of the inner sum equals $p-1$, and since there are $p$ possible values for the 
integer $s$ the probability here is $\frac{p(p-1)^2}{p^4}$.\\\\
$s\not= p$ and $0\not=a=kp$: In this setting, the inner sum equals the trace of the element $\zeta_p^k$ and is
therefore equal to $-1$. Again, there are $p$ possible values for $s$ and $p-1$ values for the integer $k$, so that in summary this
probability equals $\frac{p(p-1)}{p^4}$.\\\\
Finally, the sum of these probabilities leads to the statement of the Lemma.\hfill$\Box$\\
\vspace{0.5cm}
\begin{la}\label{lepro}
The probability ${\rm Pr}(\sigma^\prime)$, that in step {\rm (iv)} the integer $\sigma^\prime$ is recorded, satisfies 
\begin{equation}
{\rm Pr}(\sigma^\prime)=\left(1-\frac{1}{p}\right)\left(\frac{\Gamma_p^\sigma}{p}\right)^2,
\end{equation}
where $\sigma:=j_p(\sigma^\prime)|_{\Q(\Gamma_p)}$.
\end{la}
\vspace{0.5cm}
{\bf Proof.} Let $w$ be an integer, with $(w-1,p)=1$ and $w^{p-1}\equiv 1\,\,{\rm mod}\,p^2$. Then any integer 
$k\not\equiv 0\,\,{\rm mod}\,p$ can be written in the form
\begin{equation}
k\equiv w^{d_k}(1-q_p(k)p)\,\,{\rm mod}\,p^2,
\end{equation}
for some integer $d_k$. Now suppose that at the end of step (iii), we obtain a state $|a\rangle |s\rangle$, with 
$a\not\equiv 0\,\,{\rm mod}\,p$ and $s\not=p$. It then follows that the inner sum of equation (\ref{probas}) equals
\begin{equation}
\sum_{j=1}^{p-1}\zeta_{p^2}^{w^{d_a}(1-q_p(a)p)w^j(1-sp)}
=\sum_{j=1}^{p-1}\zeta_{p^2}^{w^{j+d_a}(1-(q_p(a)+s)p)}=\Gamma_p^{\sigma},
\end{equation}
with $\sigma:=j_p(q_p(a)+s)|_{\Q(\Gamma_p)}$, by definition of the element $\Gamma_p$. Since there are $p(p-1)$ ways which lead to
the same $\sigma$, the statement of the Lemma follows.
\hfill$\Box$ 
\section{Proof of the Main Theorem}
To state the proof of Theorem \ref{tmmaxgammabqp}, we first define the {\em Mirimanoff polynomial}
\begin{equation}
\gamma_p(t):=\sum_{j=1}^{p-1}\frac{t^j}{j}.
\end{equation}
This polynomial is closely related to the Fermat quotient, since
\begin{equation}\label{mpolyr1}
\gamma_p(t)
\equiv \frac{1-t^p-(1-t)^p}{p}
\equiv (t-1)q_p(t-1)-tq_p(t)\mbox{ mod }p,
\end{equation}
and therefore 
\begin{equation}\label{kpgp}
\kappa_p=\min\{n>0\,|\,\gamma_p(n)\not\equiv 0\mbox{ mod }p\}.
\end{equation}
For an introduction Mirimanoff polynomials and their basic properties, we refer to \cite{grohmannzerosfq}.\\\\
If we denote the zeros of $\gamma_p$ modulo $p$ by $\eta_{p}$ it can be shown that:
\begin{tm}\label{kbou}
There exist positive constants $c_1$ and $ c_2$ such that for all primes $p$
\begin{equation}
\kappa_p^2 < c_1 \eta_{p} < c_2 \Gamma_{{\rm max},p}.
\end{equation}
\end{tm}
{\bf Proof.} The first inequality is given by Theorem 1 in \cite{grohmannzerosfq}, while the second is shown in \cite{mydiss}, Prop. 3.16.
\hfill$\Box$\\\\
Now, in order to prove Theorem \ref{tmmaxgammabqp}, we look at the following statement:
\begin{st}\label{sta1}
There exist positive integers $s,p_0\in\N$ such that, for all primes $p>p_0$,
\begin{equation}
\Gamma_{{\rm max},p}>\frac{p}{(\log p)^s}.
\end{equation}
\end{st}
It now follows immidiately from Theorem \ref{kbou} that Conjecture \ref{cekappap} is true, if Statement \ref{sta1} is false. So, from now on, we will assume that Statement \ref{sta1} is true. From this, Lemma \ref{lepro} gives us
\begin{equation}
{\rm Pr}(\sigma_{{\rm max}}^\prime)=\left(1-\frac{1}{p}\right)\left(\frac{\Gamma_{{\rm max},p}}{p}\right)^2>\left(1-\frac{1}{p}\right)\frac{1}{(\log p)^{2s}}.
\end{equation}
Further, we define for each $\varphi\in G(\Q(\Gamma_p)/\Q)$ the real number $\alpha_{\varphi}$ by $|\Gamma_p^\varphi|=\alpha_\varphi \Gamma_{{\rm max},p}$ and (again) $\varphi^{\prime}$ by $\varphi=j_p(\varphi^\prime)|_{\Q(\Gamma_p)}$.\\\\
Finally, calling the algorithm $(\log p)^k$ times, where $k>\max\{t+10,12+4s\}$, and demanding that the difference of the expected values of $\sigma_{{\rm max}}^\prime$ and $\varphi^\prime$ lie above a certain bound,
\begin{equation}
(\log p)^k\left(1-\frac{1}{p}\right)\left(\frac{\Gamma_{{\rm max},p}}{p}\right)^2-
(\log p)^k\left(1-\frac{1}{p}\right)\alpha_\varphi^2\left(\frac{\Gamma_{{\rm max},p}}{p}\right)^2>(\log p)^{5+k/2},
\end{equation}
it follows that
\begin{equation}
\alpha_\varphi
<1-\frac{1}{(\log p)^{k-10}}.
\end{equation}
This completes the proof of Theorem \ref{tmmaxgammabqp}.

\end{document}